\documentclass[aps,twocolumn,nofootinbib]{revtex4}

\usepackage{graphicx}
\usepackage{dcolumn}
\usepackage{bm}

\usepackage{hyperref}
\hypersetup{
    colorlinks=true,
    linkcolor=blue,
    citecolor=blue,
    urlcolor=blue
}

\usepackage{graphicx}

\usepackage{titlesec}
\usepackage{amssymb,amsfonts}
\usepackage{epsfig}
\usepackage{epstopdf}
\usepackage[all]{xy}
\usepackage{amsthm}
\usepackage{dcolumn}
\usepackage{hyperref}
\usepackage{url}
\usepackage{dsfont}
\usepackage{slashed}
\usepackage{mathrsfs}

\usepackage{bbm}
\usepackage{tikz}
\usepackage{physics}
\usepackage{enumitem}
\usepackage{soul}
\usepackage{float}

\newcommand{\be}{\begin{equation}}
\newcommand{\ee}{\end{equation}}
\newcommand{\bea}{\begin{eqnarray}}

\newcommand{\eea}{\end{eqnarray}}

\newcommand*\interior[1]{\mathring{#1}}

\def\inbar{\,\vrule height1.5ex width.4pt depth0pt}
\def\IR{\relax{\rm I\kern-.18em R}}
\def\IC{\relax\hbox{$\inbar\kern-.3em{\rm C}$}}

\begin{document}

\title{A Novel Holographic Framework Preserving Reflection Positivity in dS$_d$ Spacetime}

\author{Jean-Pierre Gazeau$^{1}$\footnote{gazeau@apc.in2p3.fr}}

\author{Mariano A. del Olmo$^{2}$\footnote{marianoantonio.olmo@uva.es}}

\author{Hamed Pejhan$^{3}$\footnote{pejhan@math.bas.bg}}

\affiliation{$^1$Universit\'e de Paris, CNRS, Astroparticule et Cosmologie, F-75013 Paris, France}

\affiliation{$^2$Departamento de F\'{\i}sica Te\'orica and IMUVA, Universidad de Valladolid, E-47011, Valladolid, Spain}

\affiliation{$^3$Institute of Mathematics and Informatics, Bulgarian Academy of Sciences, Acad. G. Bonchev Str. Bl. 8, 1113, Sofia, Bulgaria}

\date{\today}

\begin{abstract}
This manuscript introduces a novel holographic correspondence in $d$-dimensional de Sitter (dS$_d$) spacetime, connecting bulk dS$_d$ scalar unitary irreducible representations (UIRs) with their counterparts at the dS$_d$ boundary ${\cal{I}}^\pm$, all while preserving reflection positivity. The proposed approach, with potential applicability to diverse dS$_d$ UIRs, is rooted in the geometry of the complex dS$_d$ spacetime and leverages the inherent properties of the (global) dS$_d$ plane waves, as defined within their designated tube domains.
\end{abstract}

\maketitle

\setcounter{equation}{0} 
\section{Introduction}
The concept of duality between a quantum theory in dS$_d$ spacetime and a Euclidean theory on its boundary was first introduced in Ref. \cite{Strominger2001}. This duality has been anticipated to capture the underlying degrees of freedom in dS$_d$ quantum gravity \cite{Strominger2001, Witten2001}. However, a crucial challenge arises as the resulting boundary theory lacks the essential property of unitarity \cite{Strominger2001} (or reflection positivity, as mentioned in Ref. \cite{Bros2002}), which is vital for a physically meaningful interpretation. Therefore, while the proposed duality construction may hold significance from a technical perspective, its direct holographic interpretation remains elusive \cite{Strominger2001, Bros2002}.

Preserving reflection positivity is a formidable challenge that extends beyond the domain of dS$_d$ quantum field theory, permeating into the realm of dS$_d$ representation theory (see Ref. \cite{Neeb}). It is essential to emphasize that elementary systems within the framework of dS$_d$ (in the Wigner sense) are intricately connected with the dS$_d$ UIRs \cite{dSBook}. Consequently, any progress in addressing this challenge within either of these interconnected contexts holds profound importance.

This manuscript particularly focuses on the realm of dS$_d$ representation theory and introduces a novel perspective that holds the potential to pave the way for ensuring reflection positivity within the dS$_d$ holographic framework. It does so by drawing on the necessity of analyticity within the complex dS$_d$ spacetime and the inherent properties of (global) dS$_d$ plane waves, which are guaranteed through their proper definition within the complex dS$_d$ spacetime domains. Specifically, it focuses on the scalar \emph{principal (`massive')} UIRs of the dS$_d$ group and establishes a profound connection that preserves the inner product.\footnote{Generally, the dS$_d$ principal UIRs hold significance from the Minkowskian perspective, given their contraction towards the Poincar\'{e} massive UIRs under the null-curvature limit \cite{dSBook, Garidi, Mickelsson}.} For a given dS$_d$ principal UIR, this connection bridges the orthonormal basis of the (projective) Hilbert space hosting the UIR within the bulk of dS$_d$ with its corresponding orthonormal counterpart on a $(d-1)$-dimensional sphere $\mathbb{S}^{d-1}$. This sphere serves as the `future' boundary ${\cal I}^+$ of dS$_d$ spacetime. Furthermore, this manuscript unveils a one-to-one correspondence that conserves the inner product between the orthonormal basis of the latter Hilbert space on the $\mathbb{S}^{d-1}$ at ${\cal I}^+$ and its antipodal orthonormal counterpart on another $\mathbb{S}^{d-1}$ sphere situated at the `past' boundary ${\cal I}^-$. 

This approach opens up a realm of exciting possibilities for further exploration and deeper comprehension of holography within the intricate context of dS$_d$ spacetime.

\emph{\textbf{Convention}}: In this manuscript, we consistently utilize the natural units where $c = \hbar = 1$.

\section{Presentation of the dS$_d$ machinery}
The dS$_d$ spacetime can be conveniently visualized as a hyperboloid embedded within a $(1+d)$-dimensional Minkowski space $\mathbb{R}^{d+1}$:
\begin{align}
M_{\text{dS}_d} = \Big\{ x \in\mathbb{R}^{d+1} \;;\; (x)^2\equiv x\cdot x = \eta^{}_{\alpha\beta} x^\alpha x^\beta = -R^2 \Big\}\,,
\end{align}
where $x^\alpha$ ($\alpha,\beta = 0,\,\ldots\,,d$) refers to the corresponding Cartesian coordinates, $\eta^{}_{\alpha\beta} = \mbox{diag}(1,-1,\ldots,-1)$ to the ambient Minkowski metric, and $R$ to the (constant) radius of curvature. 

The dS$_d$ relativity group, denoted as SO$_0(1,d)$, comprises all linear transformations within $\mathbb{R}^{d+1}$ that preserve the quadratic form $(x)^2 = \eta^{}_{\alpha\beta} x^\alpha x^\beta$, possess a determinant of $1$, and do not reverse the direction of the `time' variable $x^0$.

In a unitary representation of the dS$_d$ group, the corresponding infinitesimal generators, denoted here by $M_{\alpha\beta}$, obey the commutation rules:
\begin{align}\label{algebra}
&\big[M^{}_{\alpha\beta},M^{}_{\sigma\delta}\big] \nonumber\\
& \; = - \mathrm{i} \Big( \eta^{}_{\alpha\sigma} {M^{}_{\beta\delta}} + \eta^{}_{\beta\delta} {M^{}_{\alpha\sigma}} - \eta^{}_{\alpha\delta} {M^{}_{\beta\sigma}} - \eta^{}_{\beta\sigma} {M^{}_{\alpha\delta}} \Big)\,.
\end{align}
In the context of a scalar representation carried by the Hilbert space of Klein-Gordon square-integrable functions $\phi(x)$ on the spacetime manifold $M_{\text{dS}_d}$, these generators are given by the expression $M_{\alpha\beta} = - \mathrm{i} \big(x_\alpha\partial_\beta - x_\beta\partial_\alpha\big)$ \cite{Chernikov}. The quadratic Casimir operator for this representation is defined as $Q_0 = - \frac{1}{2} M_{\alpha\beta} M^{\alpha\beta} = -R^2 \Box_{\text{dS}_d}$, where $\Box_{\text{dS}_d}$ stands for the d'Alembertian operator on dS$_d$. 

The quadratic Casimir operator $Q_0$ exhibits the property of commuting with the action of $M_{\alpha\beta}$s. Therefore, it acts like a constant on all states in a given dS$_d$ scalar UIR:
\begin{eqnarray}\label{Wave Eq. scalar}
Q_0 \phi(x) = \langle Q_0 \rangle \phi(x)\,, 
\end{eqnarray}
where $\langle Q_0 \rangle= - \tau(\tau+{d-1})$, with $\tau\in\mathbb{C}$, denotes the corresponding Casimir eigenvalues. Then, the dS$_d$ scalar UIRs can be effectively classified based on the corresponding Casimir eigenvalues, specifically, the corresponding values of $\tau$ \cite{Dixmier, Takahashi}. 

According to Dixmier \cite{Dixmier}, the dS$_d$ scalar UIRs fall into three distinct series: principal, complementary, and discrete series. Note that the discrete series is absent in odd spacetime dimensions \cite{Lipsman, Basile}. Within our study, as already pointed out, a specific focus is placed on the representations from the principal (massive) series. For the dS$_d$ principal representations, the complex parameter $\tau$ takes the form $\tau = -\frac{{d-1}}{2} - \mathrm{i}\nu$, with $\nu\in\mathbb{R}$; $\langle Q_0 \rangle = \big(\frac{d-1}{2}\big)^2+\nu^2$. It is important to note that, the scalar principal representations with $\pm\nu$ are equivalent, sharing the same Casimir eigenvalues.

The key observation here is that, in practical terms, for a given dS$_d$ scalar principal UIR, the common dense subspace within the respective Hilbert space - the support space of the UIR - is spanned by the Klein-Gordon square-integrable eigenfunctions of the quadratic Casimir operator $Q_0$ for the assumed eigenvalue $\langle Q_0 \rangle$. Consequently, Eq. \eqref{Wave Eq. scalar}, adopted for the corresponding Casimir eigenvalue or, equivalently, the corresponding $\tau$ value, plays a fundamental role as the respective `field (wave) equation' in this group-theoretical construction.

\section{The dS$_d$ plane waves}
According to Refs. \cite{GazeauPRL, Bros2pointfunc}, for a given $\tau = -\frac{{d-1}}{2} - \mathrm{i}\nu$ ($\nu\in\mathbb{R}$), the scalar principal field equation \eqref{Wave Eq. scalar} has a continuous set of simple solutions known as dS$_d$ plane waves:
\begin{eqnarray}\label{dSplwa}
\phi^{}_{\tau,\xi}(x) = \left(\frac{x\cdot\xi}{R}\right)^\tau,
\end{eqnarray}
where $\xi$ is a vector on the null-cone $C$ within $\mathbb{R}^{d+1}$, that is, $C = \big\{ \xi\in \mathbb{R}^{d+1} \; ; \; (\xi)^2 \equiv \xi\cdot\xi = \eta_{\alpha\beta}^{}\xi_{}^\alpha\xi_{}^\beta =0 \big\}$. These dS$_d$ plane waves, as functions of $\xi$ on $C$, exhibit homogeneity with degree $\tau$. Consequently, they can be entirely characterized by specifying their values along a carefully chosen curve (known as the orbital basis) $\gamma$ of $C$.

Here, it is essential to underline that, when regarding these waves as functions on the dS$_d$ manifold $M_{\text{dS}_d}$, their definition is limited to connected open subsets of $M_{\text{dS}_d}$ and not the entire manifold, due to the possibility of the `dot product' $x \cdot \xi$ being zero. Additionally, as functions on $M_{\text{dS}_d}$, these waves exhibit multiple values since $x\cdot\xi$ can also assume negative values.

To achieve a globally defined single-valued representation of these waves, they must be treated as distributions \cite{GazeauPRL, Bros2pointfunc}. Specifically, they must be considered as the boundary values of analytically continued solutions \eqref{dSplwa} into appropriate domains within the complex dS$_d$ manifold $M_{\text{dS}_d}^{(\mathbb{C})}$:
\begin{align}\label{complex dS}
M_{\text{dS}_d}^{(\mathbb{C})} \equiv \Big\{ x+ \mathrm{i} y \in {\mathbb{C}}^5 \;; (x)^2 - (y)^2 = -R^2, \, x\cdot y=0 \Big\}\,.
\end{align}
The minimal domains of analyticity, which lead to a global single-valued definition of the dS$_d$ plane waves, are found to be the forward and backward tubes of $M_{\text{dS}_d}^{(\mathbb{C})}$, respectively defined by ${\cal{T}}^\pm = \big\{ \mathbb{R}^{d+1} + \mathrm{i} \interior{V}^\pm \big\} \cap M_{\text{dS}_d}^{(\mathbb{C})}$, where the domains $\interior{V}^\pm \equiv \big\{y \in\mathbb{R}^{d+1} \;;\; (y)^2 > 0,\; y^0 \gtrless 0 \big\}$ stem from the causal structure in $\mathbb{R}^{d+1}$. Then, by taking the boundary value (in the distribution sense) of the complexified waves from the forward ${\cal{T}}^+$ or backward ${\cal{T}}^-$ tube, while $\xi$ is merely restricted to the future light-cone $C^+ = \big\{ \xi\in \mathbb{R}^{d+1} \; ; \; (\xi)^2=0,\, \xi_{}^0>0 \big\}$, we obtain the single-valued global plane wave reading of the solutions \eqref{dSplwa}:
\begin{align}\label{mmmmmmmm}
\left(\frac{x\cdot\xi}{R}\right)^\tau =\left(\frac{(x+\mathrm{i}y)\cdot\xi}{R}\right)^\tau\bigg|_{\xi\in C^+,\, y\in\interior{V}^\pm,\, y\rightarrow 0}.
\end{align}

Notably, within this framework, as long as the analyticity domain is selected appropriately, it guarantees that in the flat (Minkowskian) limit, the dS$_d$ scalar principal waves at a given point $x\in M_{\text{dS}_d}$ precisely correspond to the conventional positive-frequency Minkowskian plane waves of a particle with mass $m$ \cite{Garidi}.

In the end, for a more comprehensive and detailed discussion of the provided material, readers are referred to Ref. \cite{dSBook}.

\section{Plane waves: generating orthonormal bases for the scalar principal UIRs in the bulk of dS$_d$}
The scalar principal waves do not display square integrability under the Klein-Gordon inner product. Nevertheless, they do give rise to the Klein-Gordon square-integrable, strictly speaking, orthonormal bases of the carrier Hilbert spaces of the scalar principal UIRs.

To illustrate this point, we invoke a set of bounded global coordinates appropriate for describing a bounded version of dS$_d$ spacetime. These intrinsic coordinates, known as conformal coordinates, can be expressed as: 
\begin{align}
x = \big(R\tan\rho, R(\cos\rho)^{-1}\mathbf{u}\big)\,, 
\end{align}
where $-\frac{\pi}{2} < \rho < \frac{\pi}{2}$ and $\mathbf{u} \in \mathbb{S}^{d-1}_1$ ($\mathbb{S}^{d-1}_1$ representing the unit $(d-1)$-dimensional sphere). Notably, the coordinate $\rho$ serves as a timelike component and plays a crucial role as a conformal time parameter.

The expression $x\cdot\xi/R$ can be expressed using the conformal coordinates as:
\begin{align}\label{dopHx.xi}
\frac{x\cdot\xi}{R} = \frac{\xi^0 e^{\mathrm{i} \rho}}{2 \mathrm{i} \cos\rho}\big(1+r^2-2r (\textbf{u}\cdot\textbf{v})\big)\,, 
\end{align}
where, above, we have defined $\xi \equiv (\xi^0,\boldsymbol{\xi}) \in C^+$, with $\boldsymbol{\xi} = (\xi^1, \ldots, \xi^d) \equiv \|\boldsymbol{\xi}\| \textbf{v} \in \mathbb{R}^{d-1}$, $\|\boldsymbol{\xi}\|^2 \equiv \boldsymbol{\xi}\cdot\boldsymbol{\xi}$ and $\textbf{v} \in \mathbb{S}_1^{d-1}$, and finally $r = \mathrm{i} e^{- \mathrm{i} \rho}$. Note that, since $\xi \in C^+$, we naturally have $\xi^0>0$ and $\xi^0 = \|\boldsymbol{\xi}\|$.

Then, by employing the generating function associated with Gegenbauer polynomials $C_n^{-\tau}(x)$ (as outlined in, for instance, Ref. \cite{Hua}), it becomes readily evident that:
\begin{align}\label{dsplwgege}
\left(\frac{x\cdot\xi}{R}\right)^\tau &= \left(\frac{\xi^0e^{\mathrm{i} \rho}}{2 \mathrm{i} \cos\rho}\right)^\tau \big(1+r^2-2r (\textbf{u}\cdot\textbf{v})\big)^\tau \nonumber\\
&= \left(\frac{\xi^0e^{\mathrm{i} \rho}}{2 \mathrm{i} \cos\rho}\right)^\tau \sum_{n=0}^\infty r^n \; C_n^{-\tau}(\textbf{u}\cdot\textbf{v}) \,.
\end{align}
Recall that $\Re(\tau)<0$. This expansion lacks validity in the context of functions. This is because the generating function for the Gegenbauer polynomials (as clearly demonstrated above) exhibits convergence only when $|r|<1$. However, in our current scenario, we have $|r|= |\mathrm{i} e^{- \mathrm{i} \rho}| =1$. Nevertheless, we are able to overcome this limitation by introducing a slight imaginary shift to the angle $\rho$, i.e., $\rho \mapsto \rho - \mathrm{i}\varepsilon\; (\varepsilon>0)$. This process ensures the convergence of the expansion, effectively leading to the extension of the ambient coordinates to the backward tube ${\cal{T}}^-$.

We also have the following auxiliary relation \cite{Hua}:
\begin{align}
C_n^{-\tau}(\textbf{u}\cdot\textbf{v}) = \frac{\Gamma\big(\frac{d}{2}-1\big)}{\Gamma(-\tau)} \sum_{k=0}^{\lfloor \frac{n}{2}\rfloor} c_k \; C_{n-2k}^{\frac{d}{2}-1}(\textbf{u}\cdot\textbf{v})\,,
\end{align}
where:
\begin{align}
c_k = \frac{\big(n-2k+\frac{d}{2}-1\big) \; \Gamma\big(k-\tau-\frac{d}{2}+1\big) \; \Gamma\big(n-k-\tau\big)}{k! \; \Gamma\big(-\tau-\frac{d}{2}+1\big) \; \Gamma\big(n-k+\frac{d}{2}\big)}\,,
\end{align}
and: 
\begin{align}
C_{L=n-2k}^{\frac{d}{2}-1}(\textbf{u}\cdot\textbf{v}) =& \; \frac{2\, \pi^{\frac{d}{2}}}{\big(L+\frac{d}{2}-1\big)\; \Gamma\big(\frac{d}{2}-1\big)} \nonumber\\
&\quad \times \;\sum_{l} {\cal Y}_{Ll}^{}(\mathbf{u})\, {\cal Y}_{Ll}^\ast(\mathbf{v})\,,
\end{align}
in which ${\cal Y}_{Ll}$ stands for the hyperspherical harmonics of degree $L$ on $\mathbb{S}_1^{d-1}$ ($\mathbf{v},\mathbf{u} \in \mathbb{S}_1^{d-1}$), $l=(l_1,\ldots,l_{d-2})$, with $L\geqslant l_1 \geqslant l_2 \geqslant \ldots \geqslant | l_{d-2} | \geqslant 0$.

Pursuing the above procedure, we arrive at the following pivotal expansion:
\begin{align}\label{mmmmmmmm}
\left(\frac{x\cdot\xi}{R}\right)^\tau = 2 \pi^{\frac{d}{2}} \, (\xi^0)^\tau \sum_{Ll} \Phi_{Ll}^\tau(x) \; {\cal Y}_{Ll}^\ast(\mathbf{v})\,,
\end{align}
where:
\begin{align}\label{ufuods}
&\Phi_{Ll}^\tau(x) = \frac{\mathrm{i}^{L-\tau} \; e^{- \mathrm{i} (L-\tau)\rho}}{(2\cos\rho)^\tau} \; \frac{\Gamma(L-\tau)}{\Gamma\big(L+\frac{d}{2}\big) \; \Gamma(-\tau)}\; \nonumber\\
&\quad\times\; ^{}_2F^{}_1\big(L-\tau \,,\, -\tau - {\textstyle{\frac{d}{2}}} +1 \,;\, L + {\textstyle{\frac{d}{2}}} \,;\, -e^{-2 \mathrm{i} \rho}\big) \; {\cal Y}_{Ll}(\mathbf{u})\,.
\end{align}

The following statements hold for any $\tau$ associated with the scalar principal representations. \emph{First}, the functions $\Phi_{Ll}^\tau (x)$ are orthonormal with respect to the Klein-Gordon inner product:
\begin{align}\label{phi inner pro}
\big\langle \Phi_{Ll}^\tau(x) , \Phi_{L^{\prime}l^{\prime}}^\tau(x) \big\rangle_{\texttt{KG}} = \delta^{}_{LL^{\prime}} \delta^{}_{ll^{\prime}}\,.
\end{align}
The Klein-Gordon inner product $\langle \cdot , \cdot \rangle_{\texttt{KG}}$ is defined, up to a possible (positive) normalization constant, by:
\begin{align}
&\big\langle \Phi(x) , \Phi^\prime(x) \big\rangle_{\texttt{KG}} \nonumber\\
&\quad = \mathrm{i}\, \int_\Sigma \Big( \Phi^\ast(x) {\partial}_\gamma \Phi^\prime(x) - \Phi^\prime(x) {\partial}_\gamma \Phi^\ast(x) \Big) \; \mathrm{d} \sigma^\gamma \nonumber\\ 
&\quad \equiv \mathrm{i}\, \int_\Sigma \Phi^\ast(x)\overset{\leftrightarrow}{\partial}_\gamma \Phi^\prime(x) \; \mathrm{d} \sigma^\gamma\,,\nonumber
\end{align}
where $\Sigma$ and $\mathrm{d} \sigma^\gamma$ respectively refer to a Cauchy surface and the area element vector on it. Considering the global coordinate choice $x=x(\rho , \mathbf{u})$ that we have adopted, the Klein-Gordon inner product explicitly reads as:
\begin{align}
&\big\langle \Phi(x), \Phi^\prime(x) \big\rangle_{\texttt{KG}} \nonumber\\
&\quad = \mathrm{i}\, \mathfrak{c}^{}_{\tau} \int_{\mathbb{S}_1^{d-1},\rho =0} \Phi^\ast (\rho,\mathbf{u}) \; \overset{\leftrightarrow}{\partial}_\rho \; \Phi^\prime(\rho,\mathbf{u}) \; \mathrm{d} \mu(\mathbf{u})\,,
\end{align}
where $\mathrm{d} \mu(\mathbf{u})$ represents the invariant measure on $\mathbb{S}_1^{d-1}$ and the (positive) constant normalization factor is:
\begin{align}
\mathfrak{c}^{}_{\tau} = 2^{2\Re(\tau)} \, e^{\pi \Im(\tau)} \; \big| \Gamma(-\tau) \big|^{2}\,.
\end{align}

\emph{Second}, $\Phi_{Ll}^\tau(x)$s exhibit infinite differentiability in terms of $x\in M_{\text{dS}_d}$.

\emph{Third}, the functions $\Phi_{Ll}^\tau(x)$, by virtue of the linear independence of ${\cal Y}_{Ll}(\textbf{u})$s, serve as conventional solutions to the respective scalar-field equation \eqref{Wave Eq. scalar} when appropriately separating variables.

Consequently, for a given $\tau$ associated with the scalar principal UIRs, the carrier Hilbert space of the respective representation can be densely generated by considering the span of all finite linear combinations of the analytic, Klein-Gordon orthonormal functions $\Phi_{Ll}^\tau(x)$.

\section{Status of the boundary theory at $\cal{I}^+$}
We now proceed to examine the behavior of the boundary theory at ${\cal I}^+$, focusing specifically on the behavior of the basis elements $\Phi_{Ll}^\tau(x)$, as they are multiplied by appropriate factors, at the limit $\rho \rightarrow +\frac{\pi}{2}$. We begin by introducing the aforesaid factors:
\begin{align}
{\mathfrak{F}}_{s}^\tau \equiv&\; (2\cos\rho)^\tau \;\Gamma(-\tau) \; \Gamma\big(\tau +{\textstyle{\frac{d}{2}}}\big) \,, \nonumber\\
{\mathfrak{F}}_{p}^\tau \equiv&\; \frac{\Gamma(L+\tau+d-1)}{\Gamma(L-\tau)} = \frac{\Gamma(L+\frac{d-1}{2}-\mathrm{i}\nu)}{\Gamma(L+\frac{d-1}{2}+\mathrm{i}\nu)} \equiv e^{\mathrm{i}\omega^{}_{L\tau}}\,, \nonumber\\
{\mathfrak{F}}_{r}^\tau \equiv&\; \frac{1}{\Gamma(2\tau +d-1)} = \frac{1}{\Gamma(-2\mathrm{i}\nu)} \,.
\end{align}
By construction, these factors respectively serve as a scale factor (recall that $\Re(\tau)<0$), a phase factor, and a regularization factor. It is worth noting that the regularization factor ${\mathfrak{F}}_{r}^\tau$ is pivotal in resolving the singularity that emerges due to the presence of the term $\Gamma(2\tau +d-1) = \Gamma(-2\mathrm{i}\nu)$ in the subsequent limiting procedure. Importantly, $\Gamma(-2\mathrm{i}\nu)$ becomes undefined when $\nu = 0$, which accurately characterizes a scenario for the scalar principal UIRs. Subsequently, we can establish the following asymptotic formula:
\begin{align}\label{LimitPhi 5.7}
&\lim_{\rho \rightarrow +\frac{\pi}{2}}\, \Big( {\mathfrak{F}}_{s}^\tau\,{\mathfrak{F}}_{p}^\tau\,{\mathfrak{F}}_{r}^\tau\,\Phi_{Ll}^\tau \big(x(\rho,\mathbf{u})\big) \Big) = {\cal Y}_{Ll} (\mathbf{u}) \equiv \Psi_{Ll}^{\scalebox{0.5}{(+)}\tau} (\mathbf{u}) \,.
\end{align}

The asymptotic modes $\Psi_{Ll}^{\scalebox{0.5}{(+)}\tau} (\mathbf{u})$ are infinitely differentiable with respect to $\mathbf{u} \in \mathbb{S}_1^{d-1}$ and are $\texttt{L}^2$ orthonormal:
\begin{align}
\big\langle \Psi_{Ll}^{\scalebox{0.5}{(+)}\tau}(\mathbf{u}), \Psi_{L^{\prime}l^{\prime}}^{\scalebox{0.5}{(+)}\tau}(\mathbf{u}) \big\rangle_{\texttt{L}^2} = \delta^{}_{LL^{\prime}} \delta^{}_{ll^{\prime}} \,,
\end{align}
where:
\begin{align}\label{L2 inner}
\big\langle \Psi(\mathbf{u}), \Psi^{\prime}(\mathbf{u}) \big\rangle_{\texttt{L}^2} = \int_{\mathbb{S}_1^{d-1}} \Psi^\ast (\mathbf{u}) \; \Psi^{\prime}(\mathbf{u}) \; \mathrm{d} \mu(\mathbf{u})\,.
\end{align}
Moreover, for a given $\tau$ associated with the scalar principal UIRs, $\Psi_{Ll}^{\scalebox{0.5}{(+)}\tau}(\mathbf{u})$s constitute a complete set of solutions for the respective scalar-field equation realized on the boundary $\mathbb{S}_{}^{d-1}$ (see, for instance, Refs. \cite{dSBook, Higuchi}). 

Hence, for a given $\tau$ corresponding to the scalar principal UIRs, $\Psi_{Ll}^{\scalebox{0.5}{(+)}\tau}(\mathbf{u})$s do indeed establish an orthonormal basis for the common dense subspace of the respective Hilbert space - the carrier of the UIR - on the $\mathbb{S}_{}^{d-1}$ at $\cal{I}^+$.

\emph{\textbf{Remark:}} In this Hilbert space, every mode $\Psi_{Ll}^{\scalebox{0.5}{(+)}\tau}(\mathbf{u})$ is linked to its antipodal counterpart: 
\begin{align}
\Psi_{Ll}^{\scalebox{0.5}{(+)}\tau}(\mathbf{-u}) = (-1)^L \Psi_{Ll}^{\scalebox{0.5}{(+)}\tau}(\mathbf{u})\,, 
\end{align}
and the following relationship holds:
\begin{align}
\big\langle \Psi_{Ll}^{\scalebox{0.5}{(+)}\tau}(\mathbf{u}), \Psi_{Ll}^{\scalebox{0.5}{(+)}\tau}(\mathbf{u}) \big\rangle_{\texttt{L}^2} = \big\langle \Psi_{Ll}^{\scalebox{0.5}{(+)}\tau}(\mathbf{-u}), \Psi_{Ll}^{\scalebox{0.5}{(+)}\tau}(\mathbf{-u}) \big\rangle_{\texttt{L}^2} \,.
\end{align}

\section{$\mbox{dS}_d$ plane waves and establishing a holographic correspondence between the bulk and boundary (${\cal{I}}^+$) Hilbert spaces}
An essential observation in the aforementioned context is the pivotal role played by the dS$_d$ plane waves as the kernel of a Fourier-type transformation, enabling a smooth transition, while upholding the principle of reflection positivity, from the Hilbert spaces that support the scalar principal UIRs in the bulk of dS$_d$ to their respective counterparts on the $\mathbb{S}_{}^{d-1}$ at $\cal{I}^+$, and vice versa. The explicit form of this kernel, $M_{\text{dS}_d} \times \mathbb{S}_1^{d-1} \ni (x,\mathbf{u}) \; \mapsto\; \mathfrak{K}(x,\mathbf{u})$, is as follows:
\begin{align}\label{ker1}
&\mathfrak{K}(x,\mathbf{u}) \equiv \frac{1}{2\pi^{\frac{d}{2}} \,(\xi^0)^{\tau}} \left(\frac{x\cdot\xi}{R}\right)^{\tau},
\end{align}
such that, from Eq. \eqref{mmmmmmmm}, we have:
\begin{align}\label{ker2}
\mathfrak{K}(x,\mathbf{u}) = \sum_{Ll} 
{\Phi}_{Ll}^{\tau}(x) \; \Big(\Psi_{Ll}^{\scalebox{0.5}{(+)}\tau}(\mathbf{u})\Big)^\ast.
\end{align}
The corresponding Fourier transformation then reads as:
\begin{align} \label{F1}
{\Phi}_{Ll}^{\tau}(x) &= {\big\langle \mathfrak{K}^\ast_{}(x, \cdot) , \Psi_{Ll}^{\scalebox{0.5}{(+)}\tau} \big\rangle^{}_{\texttt{L}^2}}\,, 
\end{align}
\begin{align}\label{F2}
\Psi_{Ll}^{\scalebox{0.5}{(+)}\tau}(\mathbf{u}) &= {\big\langle \mathfrak{K} (\cdot, \mathbf{u}) , {\Phi}_{Ll}^{\tau} \big\rangle^{}_{\texttt{KG}}}\,.
\end{align}

Clearly, for a given $\tau$ corresponding to the scalar principal UIRs, this Fourier transformation establishes a one-to-one correspondence between the Klein-Gordon orthonormal basis functions ${\Phi}_{Ll}^{\tau}(x)$, carrying the respective UIR in the bulk of dS$_d$, and their corresponding ${\texttt{L}^2}$ orthonormal asymptotic counterparts $\Psi_{Ll}^{\scalebox{0.5}{(+)}\tau}(\mathbf{u})$ on the $\mathbb{S}_{}^{d-1}$ at $\cal{I}^+$. This correspondence not only upholds the principle of reflection positivity but also ensures a clear and direct mapping.

\section{Status of the boundary theory at $\cal{I}^-$}
The remaining task to complete the aforementioned framework of holographic correspondence is to investigate the relationship between the $\texttt{L}^2$ orthonormal, asymptotic modes $\Psi_{Ll}^{\scalebox{0.5}{(+)}\tau}(\mathbf{u})$ on the $\mathbb{S}_{}^{d-1}$ at ${\cal{I}}^+$ and their respective twins $\Psi_{Ll}^{\scalebox{0.5}{(\,{\Large{-}}\,)}\tau}(\mathbf{u})$ on another $\mathbb{S}_{}^{d-1}$ at ${\cal{I}}^-$.

To do so, in line with Ref. \cite{GibbonsHawking}, we embrace the perspective of an observer in motion along the geodesic $h({x_\circ})$ passing through the point ${x_\circ}=(0,\ldots,0, x^d = R)$,\footnote{Note that the selection of this point is entirely arbitrary, owing to the SO$_0(1, d)$ symmetry of the dS$_d$ manifold $M_{\text{dS}_d}$.} situated in the $(x^0,x^{d})$-plane:
\begin{align}\label{geodesic}
h({x_\circ}) =\, \Big\{ x=x(t) \; ; \;  x^0 &= R\sinh{\textstyle{\frac{t}{R}}}, \nonumber\\
{\vec{x}} &\equiv (x^1,\ldots,x^{d-1})=0, \nonumber\\
x^{d} &= R\cosh{\textstyle{\frac{t}{R}}} \Big\}\,,
\end{align}
where $t\in\mathbb{R}$. The region comprised of all events in $M_{\text{dS}_d}$, which can be connected with the observer through the reception and emission of light signals, is defined as:
\begin{align}
{\mathfrak{R}}^{}_{h({x_\circ})} = \Big\{ x \in M_{\text{dS}_d} \; ; \; x^{d} > |x^0| \Big\}\,. 
\end{align}
This region is bounded by two distinct boundaries:
\begin{align}
{\mathfrak{H}}^{\pm}_{h({x_\circ})} = \Big\{ x \in M_{\text{dS}_d} \; ; \; x^0 = \pm x^{d}, \; x^{d}>0 \Big\}\,. 
\end{align}
which are respectively referred to as the `future horizon'/`past horizon' of the observer following the geodesic ${h({x_\circ})}$.

The parameter $t$ in the representation \eqref{geodesic} corresponds to the proper time experienced by the observer located on the geodesic ${h({x_\circ})}$. We can therefore label the `time-translation group relative to ${h({x_\circ})}$' as the one-parameter subgroup ${\mathfrak{T}}^{}_{h({x_\circ})}$ $\big(\sim \mathrm{SO}_0(1,1)\big)$ of the dS$_d$ group. The transformations associated with ${\mathfrak{T}}^{}_{h({x_\circ})}$ are hyperbolic rotations occurring parallel to the $(x^0, x^{d})$-plane. The action of ${\mathfrak{T}}^{}_{h({x_\circ})}$ on the domain ${\mathfrak{R}}^{}_{h({x_\circ})}$ is defined as follows; let $x=x(\boldsymbol{t}, \underline{\vec{x}})$ denote an arbitrary point in ${\mathfrak{R}}^{}_{h({x_\circ})}$: 
\begin{eqnarray}\label{coordinate KMS}
x(\boldsymbol{t},{\underline{\vec{x}}}) =
\left \{ \begin{array}{rl} x^0 &= \sqrt{R^2-(\underline{\vec{x}})^2} \; \sinh\frac{\boldsymbol{t}}{R}\,, \vspace{2mm}\\
\vspace{2mm} \underline{\vec{x}} &= (x^1,\ldots,x^{d-1})\,,\\
\vspace{2mm} x^{d} &= \sqrt{R^2-(\underline{\vec{x}})^2} \; \cosh\frac{\boldsymbol{t}}{R}\,, \end{array}\right.
\end{eqnarray}
where $\boldsymbol{t}\in\mathbb{R}$ and $(\underline{\vec{x}})^2=({x}^1)^2+\ldots+({x}^{d-1})^2 < R^2$. The action of $\mathfrak{T}^{}_{h({x_\circ})}(t)$, with $t\in\mathbb{R}$, on $x(\boldsymbol{t},\underline{\vec{x}})$ defines a group of isometric automorphisms of the domain ${\mathfrak{R}}^{}_{h({x_\circ})}$. It is given by $\mathfrak{T}_{h({x_\circ})}(t) \diamond x(\boldsymbol{t},\underline{\vec{x}}) = x(t+\boldsymbol{t},\underline{\vec{x}}) \equiv x^t_{}$. The associated orbits, denoted by $h_{\underline{\vec{x}}}({x_\circ})$, distinctly represent all branches of hyperbolas within the domain ${\cal{R}}^{}_{h({x_\circ})}$, which lie in two-dimensional plane sections parallel to the $(x^0,x^{d})$-plane. Note that among the given set of orbits of $\mathfrak{T}_{h({x_\circ})}$, the only orbit that represents a geodesic of ${M}_{\text{dS}_d}$ is $h({x_\circ}) \equiv h_{\underline{\vec{0}}}({x_\circ})$. [It is indeed the only orbit that extends from the `past' boundary ${\cal{I}}^-$ to the `future' boundary ${\cal{I}}^+$ of dS$_d$ spacetime.] Thus, the interpretation of the group $\mathfrak{T}_{h({x_\circ})}$ as time translation is relevant primarily for observers moving on or in the vicinity of $h({x_\circ})$, where the proximity is considered to be small compared to the radius of curvature of the dS$_d$ hyperboloid.

In this context, a remarkable phenomenon comes to light as we direct our focus toward the complex orbits of $\mathfrak{T}_{h({x_\circ})}$, referred to as $h^{(\mathbb{C})}_{\underline{\vec{x}}}({x_\circ}) = \big\{ z^t_{} \equiv z(t+\boldsymbol{t},\underline{\vec{x}}),\; t\in\mathbb{C} \big\}$. Intriguingly, all nonreal points associated with the complex hyperbolas $h^{(\mathbb{C})}_{\underline{\vec{x}}}({x_\circ})$ lie within ${\cal{T}}^\pm$, the very domains where the dS$_d$ plane waves demonstrate their analytical properties. Consequently, owing to the inherent analytic nature of the dS$_d$ waves, a significant link arises in our framework between the domain ${\mathfrak{R}}^{}_{h({x_\circ})}$ and its antipodal region: 
\begin{align}
&{\mathfrak{R}}^{}_{h(-{x_\circ})} \nonumber\\
&\quad = \Big\{ x=(x^0,\underline{\vec{x}},x^{d}) \in {M}_{\text{dS}_d}\, ; (-x^0,\underline{\vec{x}},-x^{d}) \in {\mathfrak{R}}^{}_{h({x_\circ})} \Big\}\,,
\end{align}
through the process of analytic continuation (it suffices to consider $\Im(t)=\pi$). Note that the natural time variable relevant to an observer traversing the geodesic $h(-{x_\circ})$ (antipodal to $h({x_\circ})$) is $-t$. [See Ref. \cite{Bros2pointfunc}, for more details.]

Given the freedom of selection of  the point $x_\circ \in M_{\text{dS}_d}$, as far as physics at $\mathcal{I}^\pm$ is concerned, this distinctive property inherent in our framework introduces a crucial implication: any given mode $\Psi_{Ll}^{\scalebox{0.5}{(+)}\tau} (\mathbf{u})$ on the $\mathbb{S}_{}^{d-1}$ at $\mathcal{I}^+$ is intrinsically linked to its antipodal counterpart $\Psi_{Ll}^{\scalebox{0.5} {(\,{\Large{-}}\,)}\tau}(-\mathbf{u})$ on another $\mathbb{S}_{}^{d-1}$ at $\mathcal{I}^-$, possibly up to a phase factor, as:
\begin{align} \label{+=-}
\Psi_{Ll}^{\scalebox{0.5} {(\,{\Large{-}}\,)}\tau}(-\mathbf{u}) = \Psi_{Ll}^{\scalebox{0.5}{(+)}\tau}(\mathbf{u})\,. 
\end{align}
Essentially, for a given $\tau$ corresponding to the scalar principal UIRs, this identity establishes a one-to-one mapping, while maintaining the principle of reflection positivity, between the orthonormal basis of the carrier Hilbert space of the UIR at ${\cal{I}}^+$ and its counterpart at ${\cal{I}}^-$.

\subsection{Precision on Eq. \eqref{+=-}}
To derive Eq. \eqref{+=-}, an alternative approach involves considering a limiting procedure directly, as illustrated by Eq. \eqref{LimitPhi 5.7}:
\begin{eqnarray}
\lim_{\rho \rightarrow -\frac{\pi}{2}}\, \Big( {\mathfrak{F}}_{s}^\tau\,{\mathfrak{F}}_{p}^\tau\,{\mathfrak{F}}_{r}^\tau\,\Phi_{Ll}^\tau \big(x(\rho,\mathbf{u})\big) \Big) &=& e^{-\mathrm{i}\pi\tau} \,(-1)^L {\cal Y}_{Ll} (\mathbf{u}) \nonumber\\
&=& e^{-\mathrm{i}\pi\tau}\, {\cal Y}_{Ll} (\mathbf{-u}) \nonumber\\
&\equiv& e^{-\mathrm{i}\pi\tau}\, \Psi_{Ll}^{\scalebox{0.5} {(\,{\Large{-}}\,)}\tau}(-\mathbf{u}) \,. \quad\quad
\end{eqnarray}
A straightforward comparison between this result and Eq. \eqref{LimitPhi 5.7} immediately establishes the identity \eqref{+=-}. However, it is important to note that while the alternative approach provides a more direct proof of the desired identity \eqref{+=-}, the comprehensive nature of the former approach carries significant importance. It illuminates the antipodal asymmetry described in \eqref{+=-} as a manifestation of the analytical nature of the dS$_d$ waves within the complex manifold of dS$_d$.

\section{Conclusion and outlook}
The literature on dS$_d$/CFT$_{d-1}$ has recently advanced from a more explicit application of the group theoretic features of the dS$_d$ group and sharper considerations of unitarity, as evidenced in Refs. \cite{Sun2021, Hogervorst2023, Penedones2023}. This paper contributes to this evolving literature by establishing a seamless holographic connection between the Hilbert spaces supporting the dS$_d$ scalar principal (massive) UIRs located within the bulk of dS$_d$ and their corresponding counterparts at ${\cal{I}}^\pm$, all while maintaining the principle of reflection positivity. The robustness of this framework is exemplified by three pivotal identities, identified as Eqs. \eqref{F1}, \eqref{F2}, and \eqref{+=-}. These identities offer a powerful tool to explicitly encode the physical essence of dS$_d$ within its bulk and interpret it within the context of its associated boundary theory, embodying the core principles of holography. The bedrock of this correspondence rests upon the application of the dS$_d$ plane waves, facilitating a seamless transition between these two manifestations by virtue of their analytical continuation into the appropriate domains of the complex dS$_d$ spacetime. 

While a substantial amount of work still remains ahead, we assert with strong conviction that placing reliance on this holographic framework holds the promise of establishing a comprehensive holographic correspondence on dS$_d$. Last but certainly not least, it stands out as one of the exceptionally rare constructs with the tangible potential to directly meet the minimal requirements of such correspondence, making it highly deserving of a thorough examination.

In this context, a significant objective emerges for future research, particularly concerning holography and the dS$_d$/CFT$_{d-1}$ model for quantum gravity. To address this, it becomes imperative to extend this holographic framework to encompass various dS$_d$ UIRs, with a specific focus on those associated with the discrete series in even spacetime dimensions, as exemplified in Ref. \cite{Anninos2019}. Notably, the explicit plane wave formulations of the dS$_{d=4}$ principal and discrete spin-$\frac{1}{2}$ fields, as presented in Ref. \cite{6Massive/Massless 1/2}, the principal and discrete spin-$1$ fields in Refs. \cite{6Massive 1, 6Massless 1}, the principal spin-$\frac{3}{2}$ field in Ref. \cite{6Massive 3/2}, the principal and the discrete spin-$2$ fields in Refs. \cite{6Massive 2, 6Pmassless 2, 6Massless 2} collectively pave the way forward.

Moreover, a genuine dS$_d$ holographic framework is naturally expected to provide a clear explanation for the entropy on the dS$_d$ horizon—a critical criterion in its own right. Addressing this demand presents another significant objective in the further development of the aforementioned framework. It requires the incorporation of tools and concepts from quantum information theory, including entanglement entropy and mutual information.

\section*{Akcnowledgment}
Hamed Pejhan is supported by the Bulgarian Ministry of Education and Science, Scientific Programme `Enhancing the Research Capacity in Mathematical Sciences (PIKOM)', No. DO1-67/05.05.2022. 


\end{document}